\documentclass[english,journal=jcpa,manuscript=article]{achemso}
\usepackage[T1]{fontenc}
\usepackage[latin9]{inputenc}
\usepackage{textcomp}
\usepackage{amssymb}
\usepackage{graphicx}
\usepackage{multirow}

\usepackage{babel}

\author{F. A. Gianturco}
\affiliation[innsb]
{Institut fur Ionenphysik und Angewandte Physik Universitatet Innsbruck, Technikerstr. 25/3, A-6020 Innsbruck, Austria}
\email{francesco.gianturco@uibk.ac.at}
\author{E. Yurtsever}
\affiliation{Department of Chemistry, Koc University, Rumelifeneriyolu, Sariyer, TR-34450, Istanbul, Turkey}
\author{M. Satta}
\affiliation{CNR-ISMN and Department of Chemistry, The University of Rome Sapienza, P.le A. Moro 5, 00185 Rome, Italy}
\author{R. Wester}
\affiliation[innsb]
{Institut fur Ionenphysik und Angewandte Physik Universitatet Innsbruck, Technikerstr. 25/3, A-6020 Innsbruck, Austria}

\title{Modeling ionic reactions at interstellar temperatures: the case of NH$_2^-$ + H$_2$ $\Longleftrightarrow$ NH$_3$ + H$^-$}


\begin{document}

\begin{abstract}
We present in this paper the main structural features and enthalpy details for the energy profiles of the title reactions, both for the exothermic (forward) path to NH$_{3}$ formation and for the endothermic (reverse) reaction to NH$_{2}^{-}$ formation. Both systems have relevance for the nitrogen chemistry in the interstellar medium (ISM). They are also helpful to document the possible role of H$^{-}$ in molecular clouds at temperatures well below room temperature. The structural calculations are carried out using ab initio methods and are  further employed to obtain the reaction rates down to the interstellar temperatures detected in earlier experiments. The reaction rates are obtained from the computed Minimum Energy Path (MEP) using the Variational Transition State Theory (VTST) approach. The results indicate very good accord with the experiments at room temperature, while the measured low temperature data down to 8 K are well described once we analyse in detail the physics of the reactions and modify accordingly the VTST approach. This is done by employing a T-dependent scaling, from room temperature conditions down to the lower ISM temperatures, which acknowledges the non-canonical behavior of the fast, barrierless exothermic reaction. This feature was also suggested in the earlier work discussed below in our main text. The physical reasons for the experimental behavior, and the need for improving on the VTST method when used away from room temperatures, are discussed in detail.
\end{abstract}

\section{Introduction}

The observation of molecules which belong to the Nitrogen-bearing
species has regularly been carried out in the interstellar medium
(ISM) ever since the discovery of ammonia, the first polyatomic interstellar
species, by Cheung et al \cite{a1}. Nitrogen-containing molecular
species are also considered to be useful probes of the physics and
chemistry of the ISM since they allow to perform its analysis over
a broad range of conditions. Inversion lines of ammonia have been
serving as probes of local equilibrium temperatures in molecular clouds
\cite{key-2,key-3} and rotational lines of diazenylium (NH$_{2}^{+}$)
and its deuterated isotopologue, ND$_{2}^{+}$, also allow to be 
tested at much higher densities (n \ensuremath{\sim} 105 $cm^{-3}$)
\cite{key-4,key-5} . In the diffuse ISM, for instance, the CN absorption
lines are those which allowed the first estimate of the Cosmic Microwave Background (CMB) temperature a few years back \cite{key-6}. 
Owing to its sensitivity to Zeeman
splitting and to its hyperfine structure, CN is also a powerful tool
for measuring the line-of-sight magnetic field intensity in dense
regions \cite{key-7}. 

The correct evaluation of the reservoir of nitrogen in molecular clouds
is still controversial, but it is expected to be gaseous, either in atomic
or molecular forms. Atomic nitrogen in the diffuse ISM is observed
through absorption lines in the UV \cite{key-8}. Searches for the
$N_{2}$ molecule in interstellar space have been failing until its
first detection in the far-ultraviolet by Knauth et al. \cite{key-9}
in absorption against the background star HD 124314. The derived column
density of N$_{2}$, $4.6\times10^{13}$ cm$^{-2}$, is several orders
of magnitude lower than that of atomic nitrogen, therefore indicating that nitrogen
is mainly in  atomic form. In terms of another N-bearing species, it was found that the total visual extinction for NH was 1.5 mag, equivalent to  $2.8\times10^{21}$
cm$^{-2}$, leading to abundances, with respect to hydrogen nuclei,
of $7.2\times10^{-5}$ and $1.6\times10^{-8}$ for atomic and molecular
nitrogen, respectively \cite{key-10}. The column density of N$_{2}$ was found to be  about one order of magnitude
higher than the predictions of Li et al. \cite{key-11} for translucent
clouds, including the rates for the photo-dissociation of N$_{2}$. Strong discrepancies between observations and predictions suggest
that nitrogen chemistry in such diffuse/translucent environments remains
poorly understood. 

In dense molecular clouds, where hydrogen
is molecular, the situation is even worse, because atomic neutral
N and molecular N$_{2}$ are not observable directly. In such clouds, N$_{2}$H$^{+}$, a direct chemical product of N$_{2}$,
has been observed by Womack et al. \cite{key-11} and Maret {\it et al.}
\cite{key-12}, who concluded that atomic N is most likely the dominant
reservoir of nitrogen. In addition, in pre-stellar cores with gas
densities \ensuremath{\sim}$10^{5}$ cm$^{-3}$, Hily-Blant et al
\cite{key-13} derived an upper limit on the gas-phase abundance of
atomic nitrogen which suggested that nitrogen may be predominantly
hidden in ices which are fixed on the coating dust grains. More recently, Daranlot et al.
\cite{key-14}   have found that gaseous nitrogen is mostly atomic in
dense clouds, while the dominant reservoir of nitrogen is in the form
of ammonia ices at the surface of dust grains. Nevertheless, as stressed
by these authors, the predicted amount of icy ammonia is larger than
observations in dark clouds and much larger than what is detected
in comets. Thus, one can safely consider that the reservoir of gaseous
(and solid) nitrogen in dark clouds still remains an open issue.

Another fundamental question involves the path along which ammonia
is formed in dense clouds. Le Bourlot \cite{key-15} suggested that
the gas-phase synthesis through the N$^{+}$+H$_{2}$ reaction, followed
by H abstractions and dissociative recombination reactions, was efficient
enough to reproduce the observed amounts. Very recently, however,
Dislaire et al.\cite{key-16} have revisited the experimental data available for
the N$^{+}$ + H$_{2}$ reaction and produced a new rate which is significantly
lower and falls below the critical value inferred by Herbst et al
\cite{key-17} to explain the observed abundances of ammonia. The
efficiency of the gas-phase synthesis of ammonia in comparison with  the hydrogenation
of atomic nitrogen on the surfaces of grains however remains an open
question (see  e.\ g.\ Refs. \cite{key-18,key-19,key-20,key-21}).
 Observational constraints on the amount of
ammonia locked into ices coating dust grains are difficult and infrequent,
because the N-H vibrational feature at 2.95 $\mu m$ is heavily
obscured by the 3 $\mu m$ deep water-ice bands. However, observations
of NH$_{3}$ ices in young star formation regions indicate that an
abundance of 5\% relative to water seems a reasonable value \cite{key-22}.
Perhaps related to the ammonia issue is a new constraint based on
the abundance ratios of nitrogen hydrides NH:NH$_{2}$:NH$_{3}$ towards
the Class 0 protostar IRAS 16293-2422, obtained with the HIFI (Heterodyne
Instrument for the Far-Infrared) instrument located on board the Herschel
satellite in the framework of the CHESS key programme \cite{key-23}.
The absorption lines arising from the low-lying rotational levels
of these hydrides lead to abundance ratios NH:NH$_{2}$:NH$_{3}$=5:1:300.
These abundance ratios could not be confirmed by calculations in dark
regions at a temperature of 10 K and a gas density of $10^{4}$/cm$^{3}$
\cite{key-24}. Finally, one should note that Herschel/HIFI observations
of the ortho and para forms of ammonia indicate an ortho-to-para ratio
of \ensuremath{\sim} 0.7 that could not be found nor explained with
the standard nitrogen chemistry \cite{key-25}. 

The present paper is devoted to the fate of nitrogen in yet another molecular form: i.e. to  aspects of the ionic chemistry  of  a specific molecular anion, NH$_{2}^{-}$ , which can
react in the ISM with one of its most abundant molecules, i.e. with
H$_{2}$. Thus, NH$_{2}^{-}$ will be considered in our study for
reactions at low temperatures with the environmental H$_{2}$, while we shall also
discuss the related example of H$^{-}$ reacting with ammonia, a case of a
slightly endothermic reverse reaction from the former exothermic one .
To simplify the labelling in our present discussion,the  title reactions will  be named in the present study as the Forward Reaction (FR)
and the Reverse Reaction(RR) paths, in the order given by eq. (1) below. 

Different examples of reactions involving H$^{-}$ have been studied earlier
by us in the context of clarifying its role in the formation of anionic carbonitriles
\cite{key-26}. Searches for NH$^{+}$ and NH$_{2}^{-}$ are, however,
difficult not only due to their expected very low abundances, but
also because their strongest transitions lie at frequencies that are
generally inaccessible to ground-based telescopes. With the launch
of Herschel \cite{key-27,key-28} and its sensitive Heterodyne Instrument for the Far-Infrared (HIFI),
which was designed to perform high-resolution observations at frequencies
480-1250 and 1410-1910 GHz, searches for the fundamental rotational
transitions of NH$^{+}$ and NH$_{2}^{-}$ became possible. Previous
searches for the latter cation, also using the Herschel-HIFI, had resulted in average
upper limits of the NH$^{+}$ abundance relative to molecular hydrogen
at 4 K to be $10^{-10}$, and $N({\rm NH}^{+})/N({\rm NH})$ to be around 4\textminus 7\% \cite{key-29}.

A recent search which has also included NH$_{2}^{-}$ \cite{key-30}
concluded that the NH$_{2}^{-}$  anion has very low abundances in
all models, not supporting an earlier tentative detection in SgrB2(M).
Given their simulated low abundances, that  work therefore suggests that this species will be very difficult
to detect in interstellar space in spite of its expected important
role for the Nitrogen chemistry in several environments.This was confirmed by recent terahertz spectroscopy experiments, which provided the first precise transition frequencies for the fundamental rotational transitions in para- and ortho-NH$_2^-$ \cite{key-38}.

We will specifically focus the present study on the  investigation, from computational first principles, of the molecular mechanisms 
which can descibe the  reaction  given by eq. (1) below, specifically analyzing its exothermic flow
which is producing ammonia, and called here  the 'forward' reaction (FR),
and also its slightly endothermic 'reverse' reaction (RR) which produces
the NH$_{2}^{-}$ molecule:

\begin{equation}
  {\rm NH}_{2}^{-}+{\rm H}_{2}\rightleftharpoons {\rm NH}_{3}+{\rm H}^{-}
  \label{eqreac}
\end{equation}

The above process has been experimentally studied earlier at room temperature in
both directions \cite{key-31}. It has also been studied later in an ion trap experiment, but this time both
at room temperature and then further down to much lower temperatures in order to reach
conditions similar to those of interest for the ISM environments, by Otto {\it et al.} \cite{key-32}.
The latter work \cite{key-32} focussed on the exothermic channel for the
NH$_{2}^{-}$+H$_{2}$ partners, a reaction path which we shall be discussing in detail in the present paper.

In the following Section we will present the computational
aspects which give us the various structural details of reactions
\ref{eqreac}, while in Section III we will employ a model treatment of the
the reaction rates which is known to be particularly effective
when exothermic, barrierless reactions are studied at room temperature.
We will further show a T-dependent scaling of specific molecular indeces in the latter method 
to extend its usage down to the observed low temperatures of Otto
et al \cite{key-32}. Our present conclusions will then be discussed
in Section IV.

\section{Ab initio analysis of the reaction features}

Calculations were carried out using a variety of post-Hartree-Fock
ab initio methods. One level of analysis employed the basis set choice
of the CCSD(T)/aug-cc-pVTZ//MP2/aug-cc-pVTZ. The geometry and energy
optimizations were done using the MP2 approach, while all single point
energies are calculated by using the CCSD(T) approach employed in
the GAUSSIAN 09  code \cite{key-34}. The zero-point-energy (ZPE) corrections
were included in all calculations done with both sets of methods. The same set of calculations were also carried out using the Density
Functional Theory (DFT) and employing the B3LYP/6-311++G{*}{*} expansion
level, as outlined in the Gaussian set of codes \cite{key-34}.

\begin{figure}[bt!]
 \includegraphics[width=15cm]{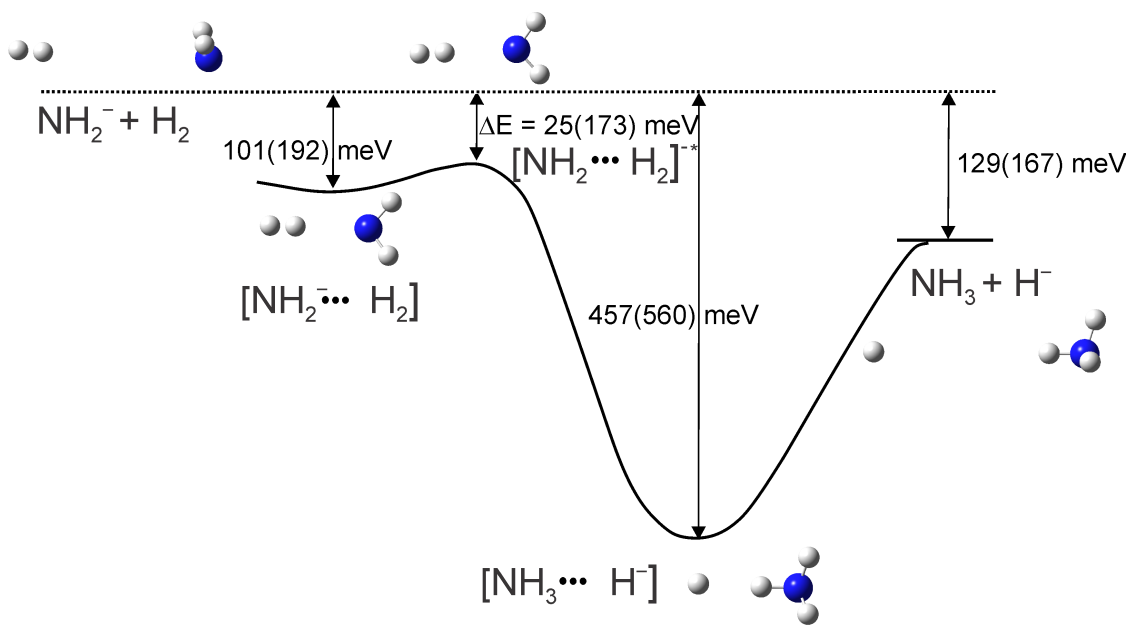}
 \caption{A schematic view of the reaction (adapted from \cite{key-32}). The energy values are from CCSD(T) calculations while those in parentheses are from DFT results.}
 \label{figschem}
\end{figure}

\begin{table}

\caption{Computed energy values and outlined geometries of the Transition State (TS), 
of the local minimum on the entrance channel (Min$_1$) and of the secondary minimum on the
exit channel (Min$_2$). Each column indicates the results from either of the
computational methods employed. The values given in brackets in the first column are those 
from \cite{key-32}. The net charges found on the reactive
hydrogen atom in the entrance and exit channels, and at the geometries
of the intermediate configurations, are given in the last set of columns
on the right-hand-side.}
\label{tabene}

\begin{tabular}{|c|c|c|c|c|c|c|c|c|}
\hline 
 & \multicolumn{2}{c|}{Energy $^a$ (meV)} & \multicolumn{2}{c|}{$R_{H-H}$ (\AA\ )} & \multicolumn{2}{c|}{$R_{N-H}$ (\AA\ )} & \multicolumn{2}{c|}{Mulliken charge on H}\tabularnewline
\hline 
\hline 
 & post-HF$^{(b)}$ & DFT$^{(c)}$ & post-HF$^{(b)}$ & DFT$^{(c)}$ & post-HF$^{(b)}$ & DFT$^{(c)}$ & post-HF$^{(b)}$ & DFT$^{(c)}$\tabularnewline
\hline 
NH$_{2}^{-}$+H$_{2}$ & 0.000 & 0.000 & 0.737 & 0.744 & 10 & 10 & 0.000 & 0.000\tabularnewline
\hline 
Min$_1$ & -101 (-100) & -192 & 0.761 & 0.755 & 2.185 & 2.064 & -0.273 & -0.238\tabularnewline
\hline 
TS & -25 (-50) & -173 & 0.926 & 0.795 & 1.475 & 1.931 & -0.121 & -0.310\tabularnewline
\hline 
Min$_2$ & -457 (-410) & -560 & 1.818 & 1.819 & 1.041 & 1.046 & -0.959 & -0.902\tabularnewline
\hline 
NH$_{3}$+H$^{-}$ & -129 (-140) & -167 & 10 & 10 & 1.013 & 1.015 & -1.000 & -1.000\tabularnewline
\hline
\end{tabular}
\\
$^{(a)}$Differences with respect to the energy of NH$_{2}^{-}$ + H$_{2}$ 

$^{(b)}$CCSD(T)/aug-cc-pVTZ//MP2/aug-cc-pVTZ 

$^{(c)}$B3LYP/6-311++g{*}{*}\\
\end{table}

A pictorial view of the various steps along the reaction, viewing the  mechanisms which play a role for the present systems are shown in figure \ref{figschem}, where the overall qualitative energy changes are marked by the computed values which are given in detail by table \ref{tabene}. It is interesting to note that the structural features reported in the table are qualitatively similar to those already reported in the earlier
work \cite{key-32} and clearly confirm the following flow of structures when
following the reaction from reagents to products along the Forward
Reaction (FR) of the exothermic process:

\begin{enumerate}

\item The exothermicity of the NH$_{2}^{-}$ reaction with H$_{2}$ varies
from the earlier value \cite{key-32} of 140 meV, given in brackets
by the 1st left column, to the present post-HF value of 129 meV and
to the 167 meV from the DFT calculations. The difference between this last value and the earlier ones can be attributed to the known property of the DFT calculations to overestimate energy gaps of reactions. It is also important 
to note here that the differences shown by the latter calculations are
additionally linked to the fact that the CCSD(T) calculations come from single-point
optimization while the DFT results follow the multidimentional optimizations along
the Minimum Energy Path (MEP) which we shall be explaiing later in this Section.

\item The location of the minimum structure in the entrance channel of the FR process (Min$_1$) is below the reference energy of the entrance channel, at an energy of -192 meV from DFT and -101 meV from CCSD(T): the earlier value from \cite{key-32} was of -100 meV.  This particular minimum
structure has an Hessian with a single negative eigenvalue so it is
 strictly only a minimum within a saddle point configuration. Its structure will be further discussed below.

 \item The transition state (TS) is located above the Min$_1$ structure.
The post-HF result gives a barrier to the latter of 76 meV (XXX not 167 meV XXX),
while the DFT calculations yields only 19 meV. These values have been
calculated along the MEP path with steps of 0.005 Angstrom and quadratic
control of convergence in the DFT procedure. In Ref. \cite{key-32}, a value of 50 meV was computed.

\item A second minimum (Min$_2$) is located at the lowest energy point along
the MEP, toward the exothermic products of the FR: it corresponds
to the H$^{-}$ bound to the NH$_{3}$ molecule and now moving toward
the final products of the Forward process. Both types of present
calculations locate it as a true minimum (no negative Hessian eigenvalues).
The CCSD(T) results are closer to the earlier calculations
of \cite{key-32}: -457 meV vs -410 meV, while the DFT results are
nearly 100 meV lower: -560 meV. From a qualitative standpoint, however, all calculations indicate similar 
orders of magnitude for the depth of this minimum structure.

\end{enumerate}

\begin{figure}[bt!]
\includegraphics[width=15cm]{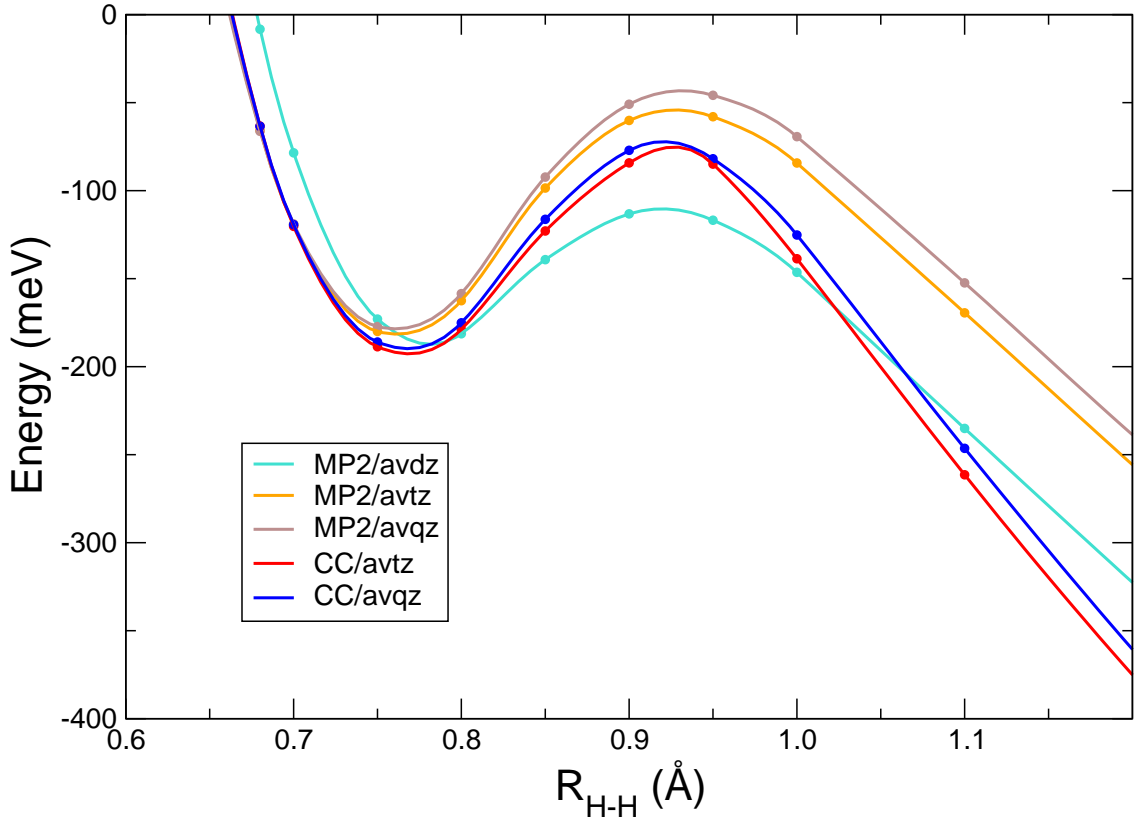}

\caption{Computational variations of the MEP reaction profile as a function of the basis set expansions from double-zed (dz) to quadruple zed (qz) and using either MP2 or CCSD levels of calculations \cite{key-34}}
\label{figmep}
\end{figure}

\begin{figure}[bt!]
\includegraphics[width=15cm]{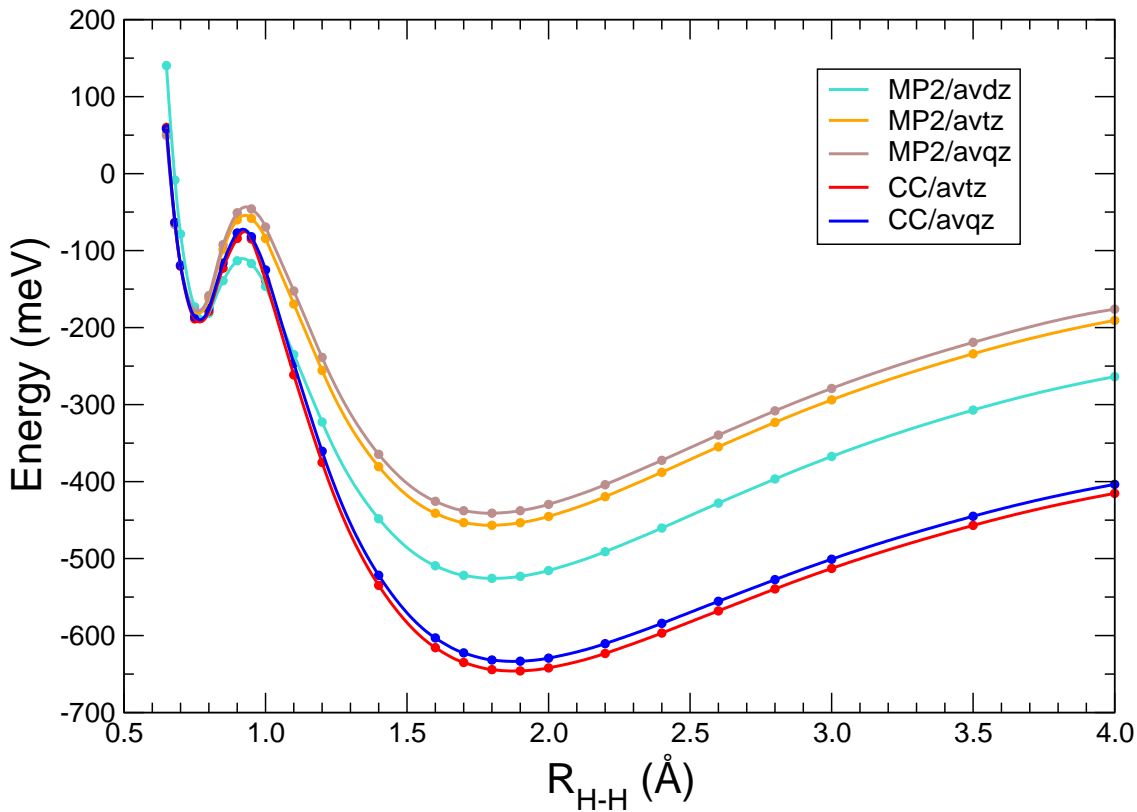}

\caption{Computed energy variations along the MEP in the region of the reagents for the exothermic reaction. The Min$_1$ and TS regions are reported. See main text for further comments.}
\label{figmep2}
\end{figure}

The potential energy curves for the reaction obtained with various methods and basis set choices are plotted in  figure \ref{figmep} to better illustrate the quality of the post-HF calculations.
A more detailed presentation in the entrance channel region of the
forward reaction, i.e. around the H-H distances of the Min$_1$ configuration,
are shown in figure \ref{figmep2}  for the same variations of basis set selection.
In general, we find the CCSD results to be more reliable for the location
of optimized minima structures than the MP2 choice, where sometime
the minima run below the asymptotic value. Thus, the CCSD results are shown in the
table \ref{tabene} in comparison with the DFT results and will be used in the
following calculations for the reaction rates and minima structures
along the reaction path.

\begin{figure}[hbt!]
\includegraphics[height=15cm]{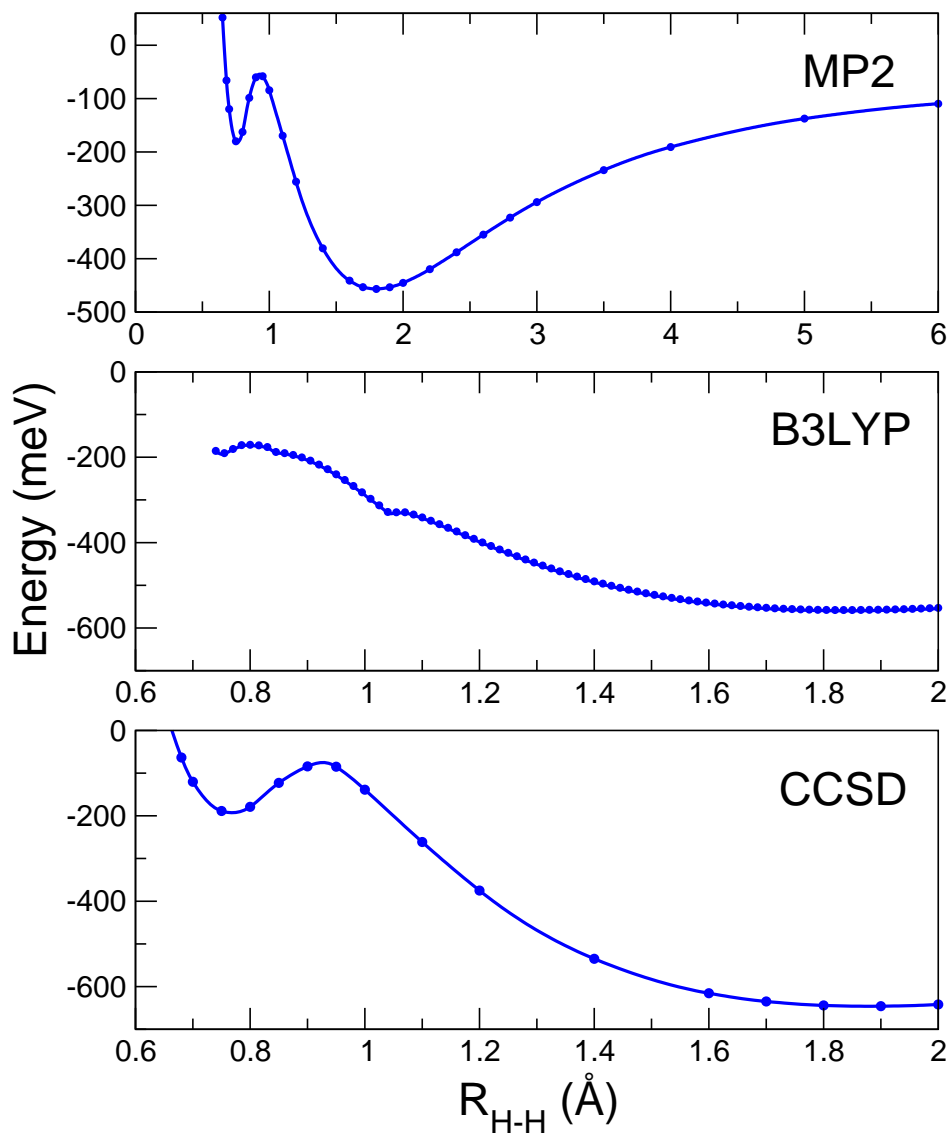}

\caption{Comparison between the computed MEP profiles . The top panel shows the fuller range of H-H distances obtained using the MP2 approach, while the lower two panels compare the CCSD(T) results(bottom panel)  with the DFT results ( middle panel)  in an enlarged view of the region near the inner minimum Min$_1$  and the small barrier to the TS when going out to exothermic products.}
\label{figmep3}
\end{figure}

\begin{figure}[bt!]
\includegraphics[width=15cm]{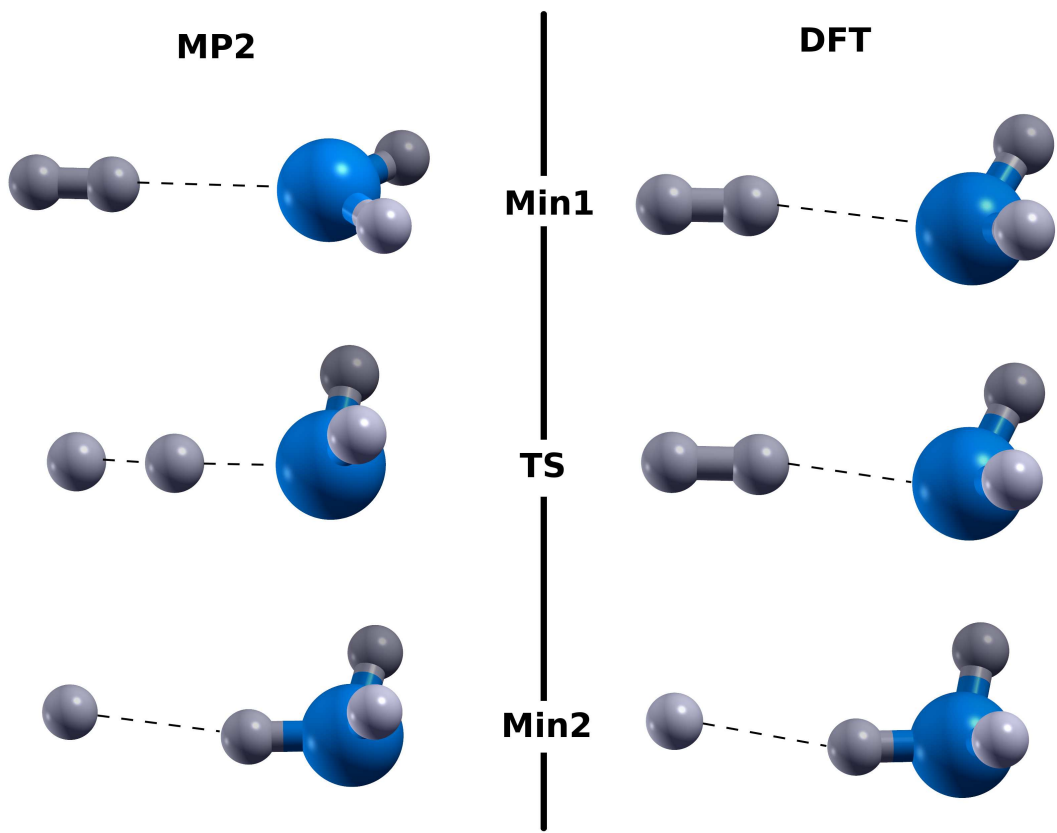}
\caption{Pictorial comparison between the significant structural details of the complex of reactants formed early on along the entrance channel, top panel, followed by the Transition State (TS) structures ( middle panel) and by the true minimum complex formed in the region of the reaction toward the products ( bottom panel). See main text for further details.}
\label{figstruct}
\end{figure}

We can make the following comments about the data shown in these figures, linking them to the numerical values for the charge migration along the reaction flow, as reported in the previous Table:

\begin{enumerate}

\item The initial minimum structure which appears at the entrance channel
of the exothermic FR process is clearly describing, in both sets of
calculations, a resonant complex where the H$_{2}$ molecule is attached
to the NH$_{2}^{-}$ reagent while some of its negative charge is
now moved onto the H-atom from the excess electron of the $\pi^{*}$
MO located on the N-atom ( see  structures in figure \ref{figstruct} ). Thus, the DFT results indicate more clearly
a charge transfer direction nearly perpendicular to the molecular
plane, which is the orientation of the excess electron MO as shown
by our earlier study \cite{key-35}. The geometries reported by both
sets of calculations are also very similar.

\item The structure of the Transition State (TS) indicates in both calculations
that the complex formed still identifies the two molecular partners
as H$_{2}$ and NH$_{2}^{-}$, although the diatomic bond is now more
stretched and the hydrogenic partner is getting closer to the anionic
partner: it could be qualitatively considered as an 'earlier' transition state. Most of the charge from the excess electron still
resides on the polyatomic partner anion while the complex reaches
the very small barrier leading to that TS structure.

\item The deepest, true minimum of the present system is clearly seen, at
the bottom of figure \ref{figmep3}, to finally form the NH$_{3}$ partner within
the complex, while the residual H$^{-}$ atom is now carrying most
of the excess negative charge from the reacting partner, thus producing
the H$^{-}$ anion: the N-H distances are now those of the final ammonia
molecule that describes the neutral molecular product.

\end{enumerate}

\begin{figure}[bt!]
\includegraphics[width=15cm]{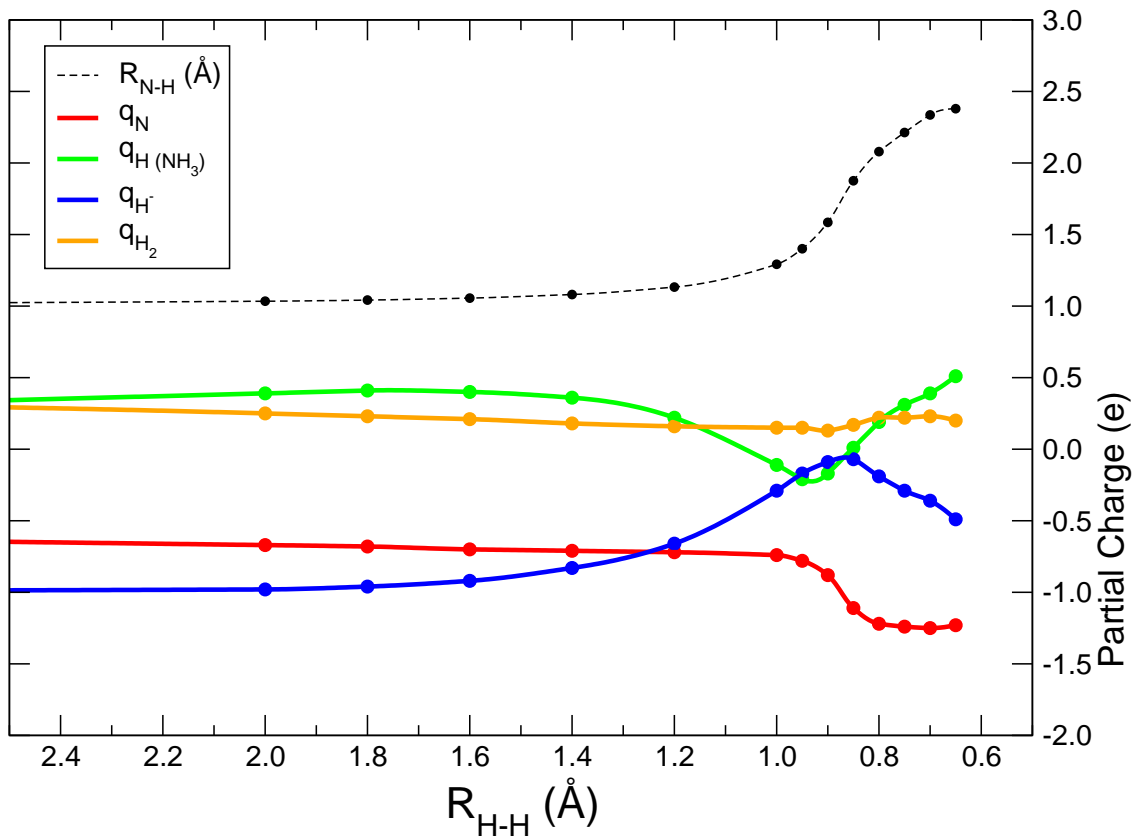}
\caption{Variation of the atomic charges as the H-H distance reaches the Charge-Transfer(CT) region during the starting of the endothermic reaction. The top curve shows the optimized distance of the N-H bond as H$^-$ approaches NH$_3$: the values in \AA\   are given on the top part of the right-hand scale.}
\label{figcharge}

\end{figure}

That the present reaction electronically involves a step where the
charge-transfer (CT) physically occurs, from the NH$_{2}^{-}$ partner
to the H$^{-}$ product, when the reacting molecules are close to
the Min$_2$ complex formation could also be gathered from the variations
of the Mulliken charges on the various participating atoms as the
reaction flows to products for the endothermic Reverse Reaction (RR), as shown by figure \ref{figcharge}.
Such changes are also confirmed by the data pictorially presented by figure
\ref{figstruct}, where we also depict the changes induced in the N-H bond of the ammonia
molecule as H$^{-}$ approaches and forms the new H$_{2}$ bond .

It is interesting to note the following features involving now  the endothermic RR
reaction mechanism: 

(i) The bond of the ammonia partner remains essentially unchanged
until the region of the CT is reached, i.e. when the H$^{-}$ distance
from the reacting hydrogen in ammonia is close to the equilibrium
bond of the H$_{2}$ product;

(ii) At that distance one sees that the extra H atom is moving away
from NH$_{3}$ to allow for the NH$_{2}^{-}$ formation, a process
also indicated by the excess charge now moving onto the N-atom, as
shown in  figure \ref{figcharge}; 

(iii) The initial negative charge on the H$^{-}$ is going to nearly
zero as the latter atom forms the neutral H$_{2}$ molecule during
the RR mechanism; 

(iv) With the same token, we also see in that figure \ref{figcharge} that the H$_{2}$
reaction partner for the FR mechanism remains always without any excess
electron charge on it, since the one on H$^{-}$ directly moves onto
the N-atom of the final NH$_{2}^{-}$ product.

\begin{figure}[hbt!]
\includegraphics[width=15cm]{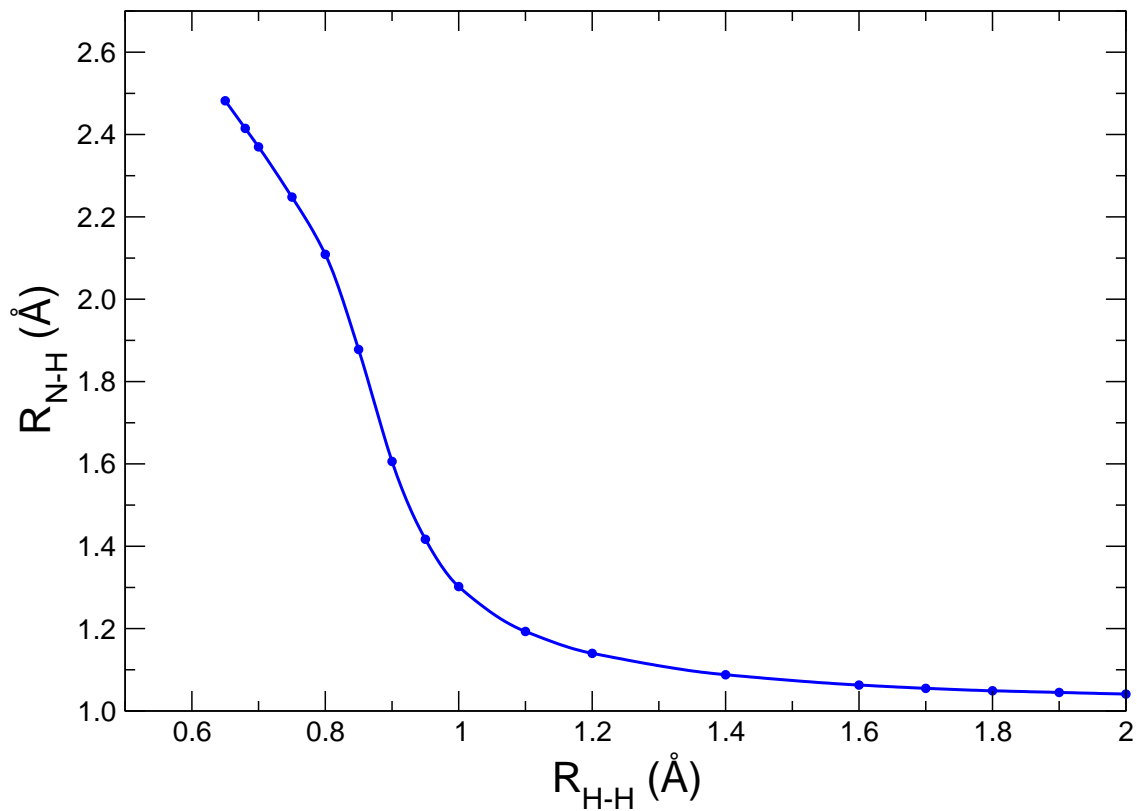}
\caption{Computed variations of the N-H distances as the H$^-$- approaches the NH$_3$ partner( from right to left) in the endothermic reaction.}
\label{figdist}

\end{figure}

An interesting piece of information regarding the acting mechanism
for the RR process could be gathered by the data presented by figure
\ref{figdist}. We report in that figure the energy variations as the N-H distance
is optimized along a series of fixed values of the H$^{-}\cdots$H
distance for the atomic anion approaching the NH$_{3}$ molecule.
What we clearly see there is that, as the H$^{-}$ gets closer to
the ammonia partner and the CT process is occurring, then the energy
optimization indicates that the reactive H-atom of the latter, now neutral
molecule, moves away from the complex to form the separated H$_{2}$
product of the RR endothermic process. Thus,it is now the N-H distance that increases
as the H$_{2}$ distance changes very little.

\begin{figure}[hbt!]
\includegraphics[width=15cm]{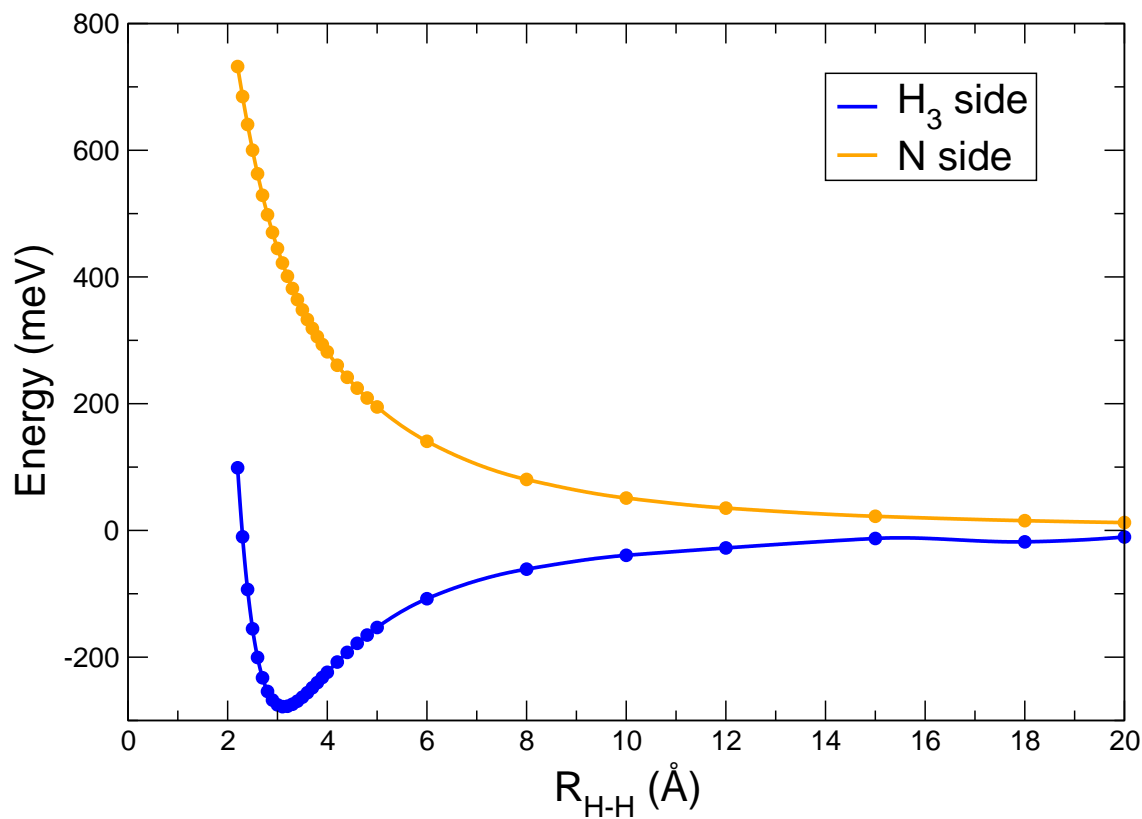}
\caption{Computed energy paths for the H$^{-}$ approach on either the N-side of the NH$_3$ partner or on the center of the (3H) face of the umbrella structure in NH$_3$. See main text for further details.}
\label{figHpath}

\end{figure}

Another structural issue for the endothermic, RR reaction, 
regards the energy efficiency of the latter as the H$^{-}$ partner
approaches ammonia from different directions. A sampling of two indicative
directions is shown by the two curves reported by figure \ref{figHpath}. In that
figure we report the minimum energy path when the H$^{-}$ approaches
the ammonia molecule along the $C_{3v}$ axis and the symmetry of
the latter is maintained. Thus, we have computed the energy optimization
steps as the anionic partner approaches from either side: the N-side
of the molecule or the face containing the three H-atoms. Here H$^{-}$
approaches along $C_{3v}$ axis. H-N distance is the variable whereas
the rest of the molecule is optimized. The (3H)-side has a minimum
around 280 meV when R(H-N) is 3.1 \AA\ while the approach on the N-side
is purely repulsive. Therefore these approaches cannot result in an H-abstraction
reaction because there is only one variable (H-N) distance and keeping $C_{3v}$
symmetry does not allow two hydrogen atoms to come together. In the
dissociative channel (where H$_{2}$ is released), we are searching
over a multidimensional grid so that we can search over a PES where two hydrogen atoms are close to
each other, as shown in the previous figures. Hence, the data in figure
\ref{figHpath} tell us that the endothermic mechanism has to involve both the CT
step and the release of one of the H-atoms of ammonia as H$_{2}$
is being formed.

\section{Modeling the reactions rates}

As we have discussed in the previous section, the energy
paths which are most likely to yield the products of both the
forward reaction (FR) and the reverse reaction (RR) involve the breaking/formation
of one of the N-H bonds in NH$_{3}$/NH$_{2}^{-}$ reagents, the formation
of one H$_{2}$ bond, and the charge-transfer (CT) step either from/to
NH$_{2}^{-}$ or from/to H$^{-}$, depending on we are considering either
the RR or the FR processes for the title system.The ensuing energy balance shows the FR to be clearly exothermic, while the RR is slightly endothermic.

\begin{figure}[hbt!]
\includegraphics[width=15cm]{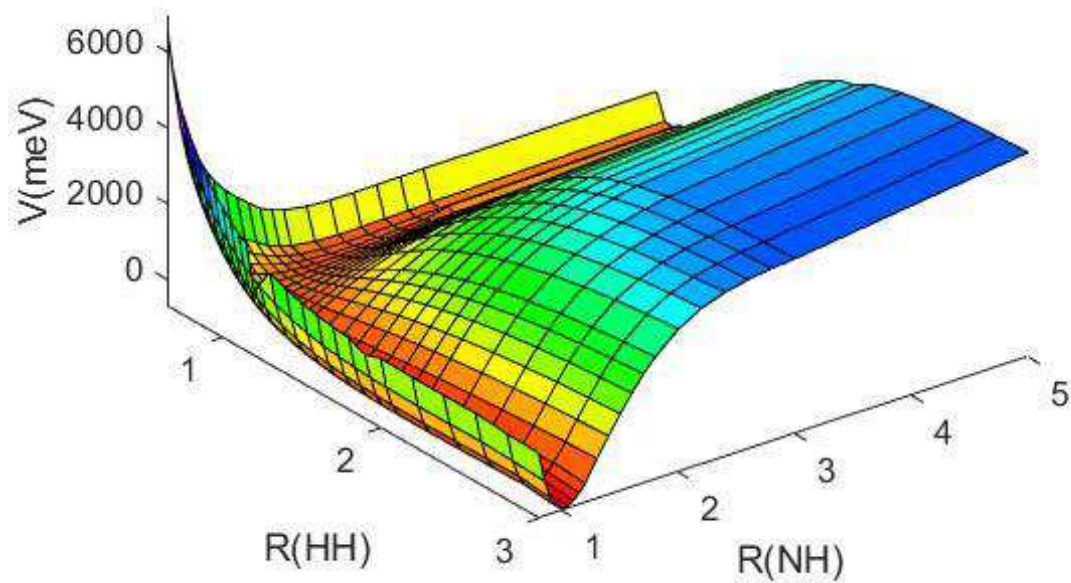}
\caption{Computed MEP energy profile for the H-H and N-H coordinates which describe the FR and RR reactions along a nearly linear approach for the two coordinates. See main text for further details.}
\label{figmeplinea}

\end{figure}

Thus, our conclusions from the calculations shown in the previous
section indicate that it is possible to treat the reactions on a reduced-dimensionality
dynamics, whereby the computed minimum energy path (MEP) is obtained
by involving chiefly two degrees of freedom along an essentially collinear,
or nearly collinear, reaction progress. A pictorial view of such path
is shown by the data in figure \ref{figmeplinea}, where we clearly see that the exothermic,
FR mechanism proceeds from the upper right of the figure down to the
lower left of it. The endothermic RR mechanism obviously follows the
opposite path.

\begin{figure}[hbt!]
\includegraphics[width=15cm]{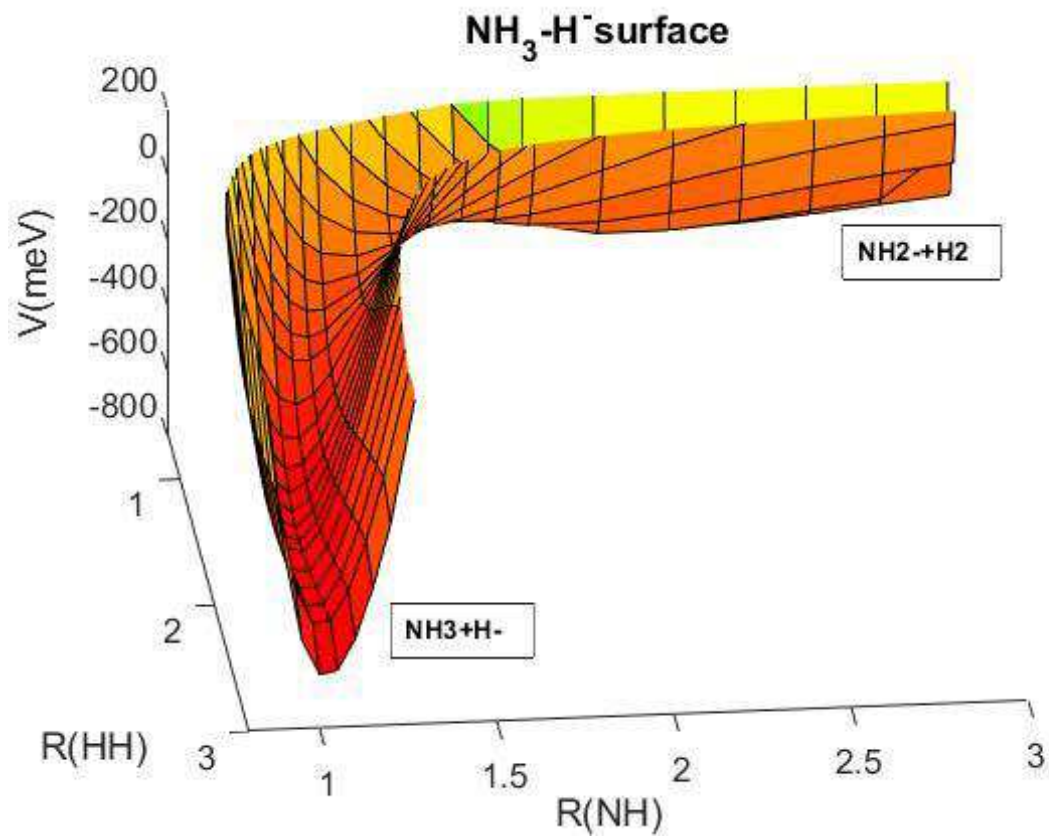}
\caption{A different view of the energy profile involved in the present MEP for the two directions of the title reaction. The labeling of the exothermic path of the FR is shown from upper right to lower left, while the endothermic path goes in the opposite direction. See main text for further details.}
\label{figmep2view}

\end{figure}

To help understand pictorially the special features of the present exothermic channel, a different view of the same MEP is presented by figure \ref{figmep2view}. Here figure \ref{figmep2view} indicates even more clearly the marked exothermicity  of the FR process and the absence of any intermediate barrier when following the NH$_{3}$-forming reaction path. We shall further analyse below the consequences of this specific behaviour of that reaction by using the Variational Transition State Theory (VTST) which we have employed before \cite{key-36} and which was already discussed in detail in our earlier work \cite{key-37}. 

The special physical features of the reaction energetics, which we have 
discussed in some detai in the previous sections, allow us to use a version of the RRKM approach based on the
Variational Transition State Theory (VTST) treatment for obtaining
temperature-dependent rates \cite{key-36} for the
case of exothermic, barrierless reactions. Just to briefly remind readers of specific aspects of this model,
we should recall that the usual TST approach searches for a stationary point along the reaction coordinate.
However, the VTST is a more general approach, which  can be applied to
reactions, such as the barrierless ones, that do not have stationary
points along their MEPs while still having a point along the MEP where
the reaction slows down, and the 'bunch of trajectories'
(to use a classical analogy) that passes over this point are the slowest
of all the possible trajectories proceeding  towards the products.
Thus, the rate coefficient is calculated by searching for the minimum of the molecular partition function $Q$
along an extensively computed MEP, while the standard Transition State Theory (TST)
calculates the rates  starting from the $Q$ of the molecular
geometry corresponding to the barrier top (stationary point) of the
reaction. In both cases one therefore needs the $Q$ of the reagents.

It is important to also note here that, in the VTST modeling, any molecular structure associated to a minimum
of its molecular partition function along the MEP can have no imaginary
frequencies: often this is the case, especially for those MEP whose energies
are always smaller than those of the reactants' entrance channels.
In other words, the TS geometry, according to the VTST
approach, is not a stationary point. Therefore in our case, where
we study the exothermic FR mechanism for the reaction: NH$_{2}^{-}$+H$_{2}$ $\rightarrow$ NH$_{3}$ + H$^{-}$,
the minimum of the $Q$s along the MEP is slightly shifted with respect
to the barrier top that leads the system towards the products region.
At this point all the frequencies are positive even if the MEP has
a negative curvature in that reaction coordinate region. This can
be explained because in the calculation of the MEP all the coordinates,
except the reaction coordinate, are simultaneously optimized. On the
contrary, in the frequency calculations for the TST model, each element of the second
derivative matrix is obtained by fixing all the coordinates except
those corresponding to the normal mode of that element.

This distinction provides a marked difference with respect to the
standard transition state theory applied to reactions with a barrier,
where the bottleneck of the 'reaction trajectories'
is always located exactly at the top of the barrier and there, almost
always, at least one of the frequencies is imaginary. But even in
that case one can calculate the ZPE since the normal mode corresponding
to the imaginary frequency is neglected there, and therefore also its molecular
partition function. This is the reason why in the usual  TST the $Q$ has a
minimum for the molecular geometry corresponding to the barrier top.
This is the difference between VTST and TST: in the former modeling one searches for
the  minima  of the $Q$s that develop all along the MEP set of structures and includes
 all the residual degrees of freedom.

We have recapped the VTST method above to show that it basically assumes the formation of a transition-state complex all along the exothermic energy path from reactants to products and it further controls the efficiency of product formation via the relative energetics between the partition functions of that complex and those of the initial reagents. Furthermore, one assumes a strong-coupling approximation whereby the degrees of freedom within the evolving TS complex are strongly coupled among themselves, but, due to the ususually low pressure conditions in experiments and in the ISM, there is no coupling  with the surrounding thermal bath provided by the other molecules in the ambient gas. One should be reminded that the collision frequencies with the other molecules would be much larger than microseconds, while the intramolecular vibrational redistribution (IVR) time scales with the non-reacting degrees of freedom would be of the order of 0.1 ps. This time scales needs to be compared with the time spent at the TS within the MEP during the reaction: this may vary from about 150 fs at 300K to about 350 fs at 10K.

From the above, it follows that, depending on temperature, the crossing of the MEP during the reaction will occur on a shorter time-scale than that of the characteristic intramolecular vibrational rearrangements. This means that, once the TS area is reached by the reagents along the exothermic MEP, the reactants are in microcanonical equilibrium with their non-reactive surroundings and also with the reactive coordinate and its kinetic and potential energy content. Under room temperature conditions the TS reactive degree of freedom can then efficiently release its energy gained during the exothermic process to the relative kinetic energy of the products while avoiding to transfer energy to the other non-reactive modes. That such assumption could in fact break down as the temperature of the reactive process becomes much smaller than room temperature will be discussed in the following.

The complex partition function, $Q_{TS}^{\ddagger}$, can then be obtained as the product of the $Q$s for the conserved mode partition functions within that complex, $Q_{cons}^{\ddagger}$, times the $Q$s of the translational mode:

\begin{equation}
  \label{eqqpart}
Q_{TS}^{\ddagger}(T)=Q_{cons}^{\ddagger}(T)Q_{trans}^{\ddagger}(T)
\end{equation}

Within the VTST the activation energy is zero and therefore the standard exponent of the Arrhenius equation is equal to unity. This now means that it will be the pre-exponential factor of the rate formulation by that equation which becomes the only significant part which needs evaluation to obtain the reaction rates $K(T)$.
The latter quantity can now be obtained as the ratio between the relevant $Q$s of the TS complex and those of the reagents, as a function of the reaction temperatures:

\begin{equation}
 \label{eqrate}
K(T)=\frac{k_{B}T}{h}\frac{Q[(NH_{2}^{-})\cdots(H_{2})]}{Q[NH_{2}^{-}]Q[H_{2}]}
\end{equation}

Here $k_{B}$ is the Boltzmann constant and $h$ is the Plank's
constant. The $Q$s of the reaction complex are variationally minimized
along all the optimized geometries of the MEP. After the variational
minimization we therefore obtain at each temperature the TS geometries
at the lowest possible energy along the MEP paths and within the VTST
model (see  e.\ g.\ \cite{key-36}). In other words, the model assumes
that the extra energy acquired by the reactive degree of freedom
of the TS, as it descends exothermically along the MEP towards the
products, is efficiently transferred into the relative kinetic energy of the products, so that at
each temperature value the reaction rate can progress within a peudo-canonical equilibrium within the complex.
It is interesting to note at this point that
the rate coefficient as obtained from the above equation is, at the lowest
temperatures, linked to the behaviour of the vibrational part of the
Qs for that specific TS geometry since all other degrees of freedom
are now \textquotedbl{}frozen\textquotedbl{} to their lowest values
in order to satisfy to the equilibrium condition required by implementing
the VTST treatment of the rates. As we descend to increasingly lower
temperatures, however, the TS's 
$Q$s relate directly to the zero-point energy (ZPE) values of the modes
under consideration since $Q_{vib}$(at low T) can be written as:
$e^{-E_{ZPE}/k_{B}T}$. Thus the overall rates at low-T would be linked
with the features of the corresponding ZPE energy values within the
$Q$s:

\begin{equation}
 \label{eqklowt}
K(low T)\propto T e^{\frac{-E_{ZPE}^{TS}+E_{ZPE}^{Reacts}}{K_{B}T}}\propto T e^{\frac{-\Delta_{ZPE}}{K_{B}T}}
\end{equation}

In the present reactions we have that $E_{ZPE}^{reacts}=E_{ZPE}^{NH_{2}^{-}}+E_{ZPE}^{H_2}$. The quantity $\Delta_{ZPE}$ thus defines the energy differences between the E$_{ZPE}$ of the reactants
and those of the TS complex. It therefore controls the slope of the
reaction rate $K$ (at low temperature) which is defined by the above equation:
a large value of that difference causes a greater decrease of $K(T)$
as $T$ decreases. The ZPE values of our present MEPs depend on: (i)
the H-H-N vibrations at the TS geometries and: (ii) the intramolecular
vibrational frequencies within the remaining atoms in the reactants
(in our case NH$_{2}^{-}$ and H$_{2}$). We shall further discuss
this point below when analyzing the low-T behavior of our calculated
rates. 

For the exothermic reaction, the FR reaction, we evaluated the minima
of the partition functions Qs of the TS complex along the computed
MEP configurations from reactants to products. For this reaction
we find that this minimum is practically coincident with the location
of the pseudo-TS along the path. However, when we did the calculations
for the endothermic RR process, that minimum is located in the region
of the MEP which is very close to the final products, as discussed
earlier in Section II. Thus, we find that the reaction bottleneck
for the FR mechanism is located just before the 'downhill'
path to products which brings the MEP to its absolute minimum. On
the other hand, for the RR that minimum of the TS would be located
at the MEP minimum as requested by the VTST model. When applying eq. (3)
to obtain the rates for the two reactions we obtain directly the rate
of the exothermic FR process, while for the endothermic path, the
RR process, we still have to multiply the l.h.s. ratio by $\exp(-H/k_{B}T)$
where $H$ is the reaction enthalpy of the endothermic process, as discussed
in Section II. This additional factor provides the main difference
of behavior between the FR and RR mechanisms at the low temperatures
of the ISM environment, as we shall further discuss below. 

\begin{figure}[hbt!]
\includegraphics[width=10cm]{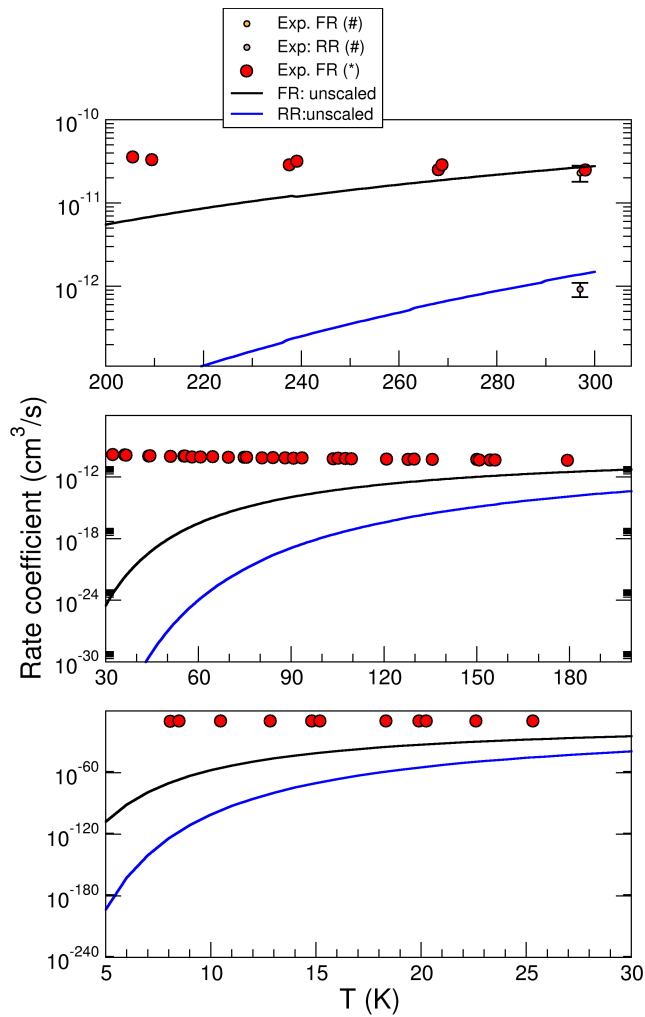}

\caption{Computed VTST rates for the exothermic (black) and endothermic (blue) paths of the title reactions. The experimental points are from: (*) (Otto et al.) \cite{key-32}  and (\#) (Bohme) \cite{key-31}. See main text for further details.}
\label{figrate}
\end{figure}

We report in the next figure \ref{figrate} the behavior of both rates for the
two directions of the title reaction. The three panels of that figure
indicate the temperature dependence of the  rates within different
ranges of T. We can easily see that the two existing measurements
at room temperature reported in the upper panel agree very well with
the present calculations for both the FR and the RR mechanism. It
therefore follows that the VTST approach, using the present ab initio
calculations, works realistically in describing this reaction in the temperature
region of the laboratory experiments. At the same time, when comparing
with the other existing experiments which were carried out below room
temperature, and down to about 8 K, we now see that the computed FR
rates decrease rapidly with T, while the experimental rates increase with decreasing T.

To try to better understand the possible physical origins of these
differences, we should begin to note first that the standard, textbook VTST approach includes
the molecular rotations through the approximate formula:

\begin{equation}
    \label{eqqrot}
Q_{rot}(T)=\frac{1}{s}\left(\frac{k_{B}T}{hc}\right)^{3}\frac{1}{2}\left(\frac{\pi}{A\times B\times C}\right)
\end{equation}

Here, $s$ identifies the multiplicity and $A$, $B$, $C$ are the rotational constants
of the TS and of the individual molecular reactants. This rather strong approximation
presumes that the temperature of interest be: $T >> hcX/k_{B}$ , where $X$ is the largest of the rotational constants. We have done the calculations at room temperature using both the above simplification and the exact quantum approach, finding only marginal differences in the calculated rates.
However, the use of the above equation becomes a limitation of the standard VTST treatment once one moves to low temperatures. For instance, in the present case, for NH$_{2}^{-}$
the value of $T$ should be $>> 32.89$ K, while in the
case of H$_{2}$ should be $>> 86.87$ K. We therefore
see that the standard VTST approach would be valid at room temperatures
but starts to fail as the temperature is lowered down to $T < 200$ K. 

If we now look at the evaluation of the present rates at low-T as given by eq. \ref{eqklowt} we see that the large values of the $\Delta_{ZPE}$ between the geometry of the TS and that of the reagents is the chief cause for the rapid dropping of K(T) as the temperature decreases. We have therefore attemped to improve on the a VTST approach by devising a scaling of such $\Delta_{ZPE}$ values beween 300 K and 8 K, using the room-temperature agreement as a reference value. Let us therefore write the above quantity in a temperature\textendash dependent form

\begin{equation}
    \label{eqdzpe}
\Delta_{ZPE}\Longrightarrow\frac{(\tau+T)}{300K}\Delta_{ZPE},
\end{equation}

where $\tau$ is a fitting parameter, which we give the value of
5.3 K. We have also replaced the approximate formulation of eq. \ref{eqrate}
for H$_{2}$ with its quantum formulation of the their rotational
partition functions \cite{key-40}. The calculations indicated by eq. \ref{eqrate}  are now
done again with the replacement of the fixed-value $\Delta_{ZPE}$ with its T-dependent
behaviour from eq. \ref{eqdzpe}. The results are shown in the Figure \ref{figratemodel}  below. 

\begin{figure}[hbt!]
\includegraphics[width=18cm]{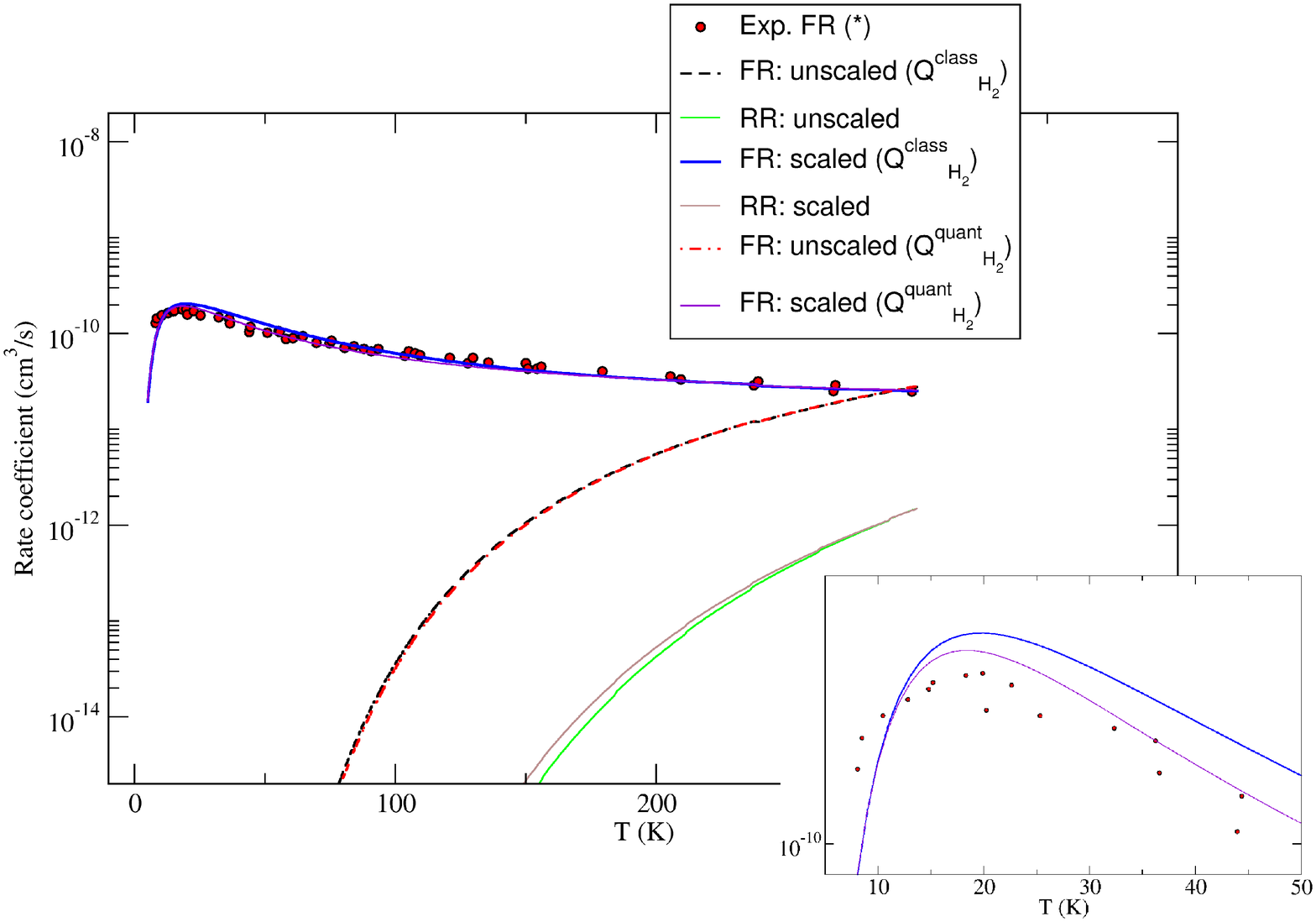}
\caption{Computed FR reaction rates using the VTST approach with and without quantum formulation of the Q$_{rot}$ (dashed red and black curves). The same treatments for the RR endothermic reaction are given by the solid green and pink curves. The same comparison but using the scaling of eq. \ref{eqdzpe}  is given by the solid blue and red curves. The bottom-right insert  reports an enlarged view of  the experimental and calculated  rates in the low temperature region.}
\label{figratemodel}
\end{figure}

It is interesting to note in that figure how the FR exothermic reaction now yields rates which agree splendidly with the experiments down to the lowest measured temperatures provided by experiments. It indicates that the T-dependent modification of the standard VTST  at low-T is essential to recover a computed behaviour that correctly describes the involved physics. 

To try to better understand such reasons, we rewrite the scaling formula in the following way:

\begin{equation}
  \label{eqtau}
e^{-\frac{\Delta_{ZPE}}{k_{B}T}}\Longrightarrow e^{-\frac{\tau}{300K}\frac{\Delta_{ZPE}}{k_{B}T}}e^{-\frac{\Delta_{ZPE}}{k_{B}300K}}=\left(e^{-\frac{\Delta_{ZPE}}{k_{B}T}}\right)^{\frac{\tau}{300K}}e^{-\frac{\Delta_{ZPE}}{k_{B}300K}}
\end{equation}

The above expression is now factored into two different terms which could be given different physical meaning and different roles in explaining the effects from this T-dependent scaling. The first term describes the same exponential as in eq. (4) which, as before, describes the low-T behavior of the rates, but is now modified by the presence of an exponentially reducing factor of ($\tau/300 {\rm K}$). The second term reports the same unchanged term as employed earlier in eq. (4), but now kept fixed at the reaction\textquoteright s room temperature of 300 K.

If we consider these two terms to describe (as a classical analogy) the total 'trajectory flux' during
reaction, then we could say that the second term gives the correct reaction flux of trajectories at room temperatures, where the TS state, which is beginning to run 'downhill' towards the products, correctly manages at the higher $T$ values to rapidly cross the reactive region of the MEP and therefore to efficiently transfer towards the products the excess energy it has gained during the exothermic run along the MEP, thereby following the variational lowest energy path discussed earlier. The surrounding bath comprises the residual vibrational modes of the reacting  NH$_{2}^{-}$  and the H$_{2}$ vibrations. On the other hand, the first term of eq. (7) is markedly reduced from its unscaled form by the introduction of the ($\tau/300 K$) factor: it describes  a small portion of the trajectories (the larger portion being given by (1-$\tau$/300K)= 0.98\%) which manages, as
$T$ decreases, to transfer the reaction enthalphy to the relative kinetic energy of  the prodcuts without heating the non reactive modes when going below room temperature, while the larger portion of the trajectories  (second term of eq. (7)) is such that the MEP crossing is so slow that within the complex there is enough time to, at least partially, transfer the excess reaction energy to the non-reactive modes.

Thus, for exothermic reactions which are investigated well below room temperatures, the TS structure is not efficiently keeping its thermal balance during the slow  MEP crossing of the reaction region, and therefore the VTST model fails at such low temperatures. Our scaling procedure is thus correcting for this failure by reducing the flux contributing to obtaining the final VTST rates to only that number which manages to keep the required pseudo-canonical equilibrium of the VTST treatment, as the temperature goes increasingly below room temperature.  

\begin{figure}[hbt!]
\includegraphics[width=10cm]{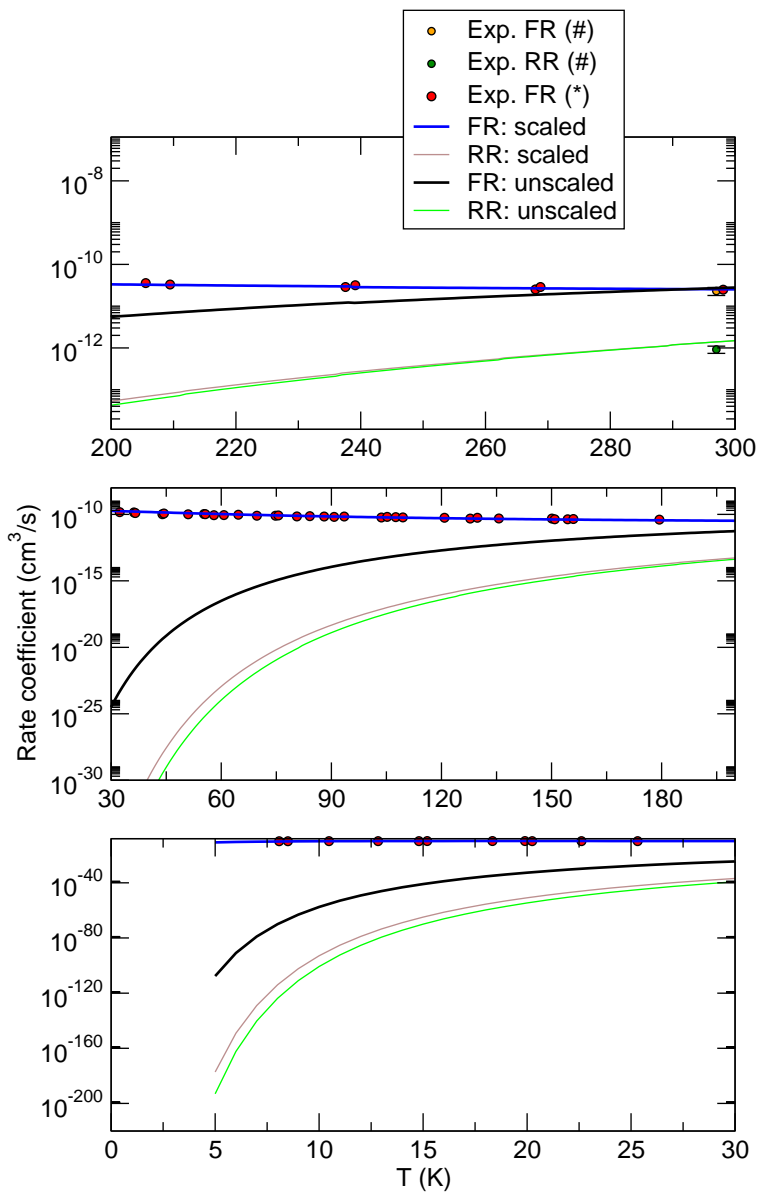}
\caption{Computed, scaled and unscaled, VTST reaction rates for the exothermic forward process and endothermic reverse process. The experimental points are from: (*) (Otto et al.) \cite{key-32}  and (\#) (Bohme) \cite{key-31}. See main text for further details.}
\label{figratescal}
\end{figure}

Such scaling correction only holds for the exothermic reactions with small or no barrier and at low temperatures. It is not sufficiently corrected for the endothermic channel, where the rapid decrease of the rates with decreasing $T$ is not substantially altered when a barrier appears in going to the products' region. Such differences in behavior are presented in the three panels of figure \ref{figratescal}. The data in the upper panel show again that, at room temperature, both FR and RR mechanism are correctly described by the present ab initio modeling of the VTST reaction rates. That panel also indicates how the unscaled treatment begins to loose agreement with experiments (only available for the exothermic path) when $T$ becomes lower than about 250 K. The next two panels further show how dramatically the standard VTST reaction rates decrease with $T$ and markedly depart from the available experimental data for the FR mechanism, clearly reaching strongly unphysical values in the bottom panel. On the other hand, we see that the empirical  reduction of the number of trajectories which maintain the VTST conditions allows now the computed rates to agree very well with experiments down to the lowest available temperature.  We should also note here that the same scaling, when applied to the endothermic reaction, does not manage to increase their values as $T$ decreases. Whether experiments in that thermal region will indicate this behaviour to be the correct, still remains to be verified by possible new measurements for the endothermic reaction.

To comply with some of the rates $T$-dependence used in astrophysical data bases, the data of the rates in figure \ref{figratescal} have been fitted with the parametric equation
$K(T)=\alpha (\frac{T}{300 {\rm K}})^\beta e^{-\gamma /T}$, and the fitted parameters are reported  in table \ref{tabrate}.

\begin{table}[hbt!]

\caption{Fitting parameters for the rate coefficients of the forward and reverse
  title reactions with and without scaling.}
\label{tabrate}

\begin{tabular}{|c|c|c|c|c|}
\hline 
 & T range & $\alpha$ ($10^{-10}cm^{3}/s$) & $\beta$ & $\gamma$ (K)\tabularnewline
\hline 
\hline 
\multirow{3}{*}{FR unscaled} & 5-100 & 9.6361 & -1.0214 & 1132.6\tabularnewline
\cline{2-5} 
 & 100-200 & 9.0036 & -0.2580 & 1039.4\tabularnewline
\cline{2-5} 
 & 200-300 & 2.8511 & 1.0591 & 697.3\tabularnewline
\hline 
\multirow{3}{*}{FR scaled} & 5-100 & 0.1615 & -1.4755 & 29.09\tabularnewline
\cline{2-5} 
 & 100-200 & 0.1985 & -0.4333 & -67.62\tabularnewline
\cline{2-5} 
 & 200-300 & 0.1304 & 0.0872 & -195.97\tabularnewline
\hline 
\multirow{3}{*}{RR unscaled} & 5-100 & 8.0267 & 1.0126 & 1885.7\tabularnewline
\cline{2-5} 
 & 100-200 & 12.9920 & 0.3823 & 2033.8\tabularnewline
\cline{2-5} 
 & 200-300 & 10.8181 & -0.6917 & 2133.8\tabularnewline
\hline 
\multirow{3}{*}{RR scaled} & 5-100 & 6.2063 & -0.5351 & 1949.0\tabularnewline
\cline{2-5} 
 & 100-200 & 4.6514 & 1.1692 & 1719.3\tabularnewline
\cline{2-5} 
 & 200-300 & 4.1581 & 1.2913 & 1690.5\tabularnewline
\hline 
\end{tabular}

\end{table}

\begin{figure}[t]
\includegraphics[width=15cm]{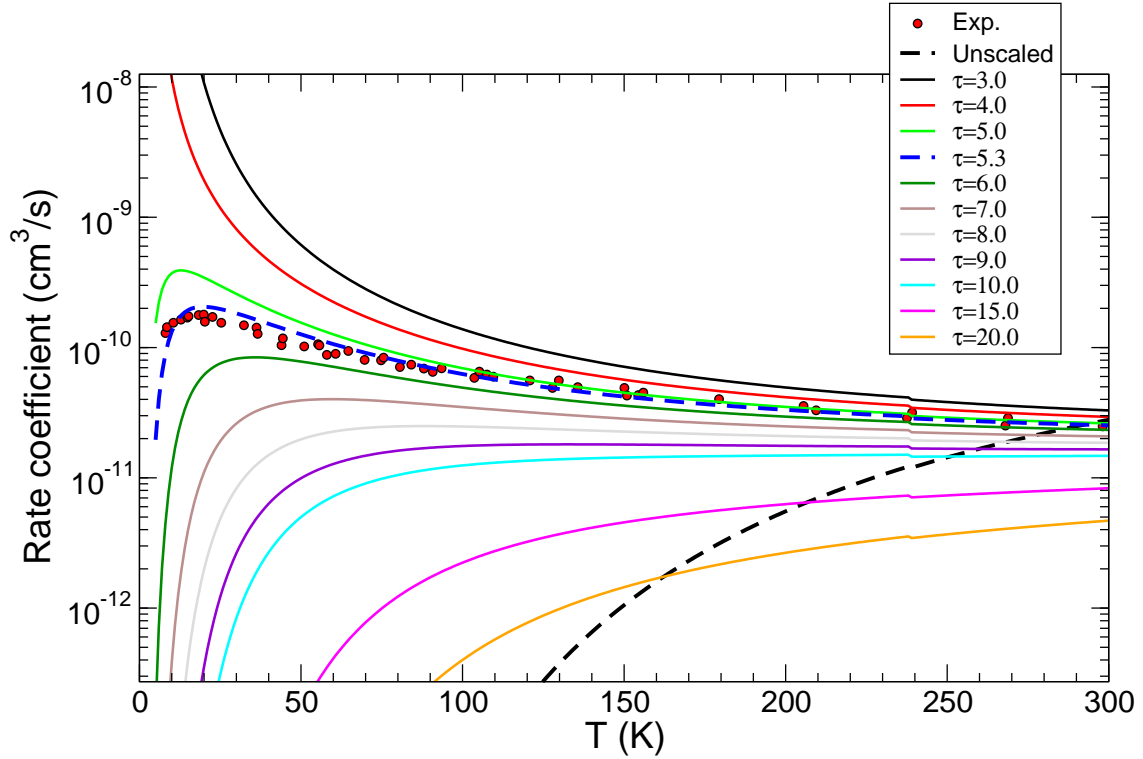}
\caption{Effect on the computed rates of changing the parameter $\tau$ over an ample range of values. The dots refer to the experimental points  from: (*)(Otto et al.) \cite{key-32} and (\#) (Bohme) \cite{key-31}. See main text for further details.}
\label{tauscan}

\end{figure}

Since the fitting parameter in eq. (6) modifies the final behaviour of our computed rates, it is of interest to try to see whether it also carries a physical meaning linked to the specific value chosen for it in the calculations. To this end, it is useful to compute its effect on the computed rates when that parameter is made to vary over a large range of values. The results of our computations are shown by figure \ref{tauscan} where the values of the parameter in eq. (6) vary from 3.0 to 20.0, showing that the accord with the experimental data only exists for parameter's values around 5.0.

It is tempting to go further in our understanding of the role of the $\tau$ parameter by linking it to a sort of effective temperature, $T_{eff}$, and to the $\Delta_ {ZPE}$ quantity computed via eq. (4):

\begin{equation}
  \label{teff}
\exp \left[-   \left(  \frac{ \tau+T}{300 {\rm K}}   \times   \frac{ \Delta_{ZPE}} {KT} \right)  \right] = \exp \left(-\frac{ \Delta_{ZPE}} {KT_{eff}} \right)
\end{equation}
where we define: 
\begin{equation}
\label{teff2}
T_{eff} = \frac{300 {\rm K} \times T} {(\tau + T)}
\end{equation}

\begin{figure}[t]
\includegraphics[width=15cm]{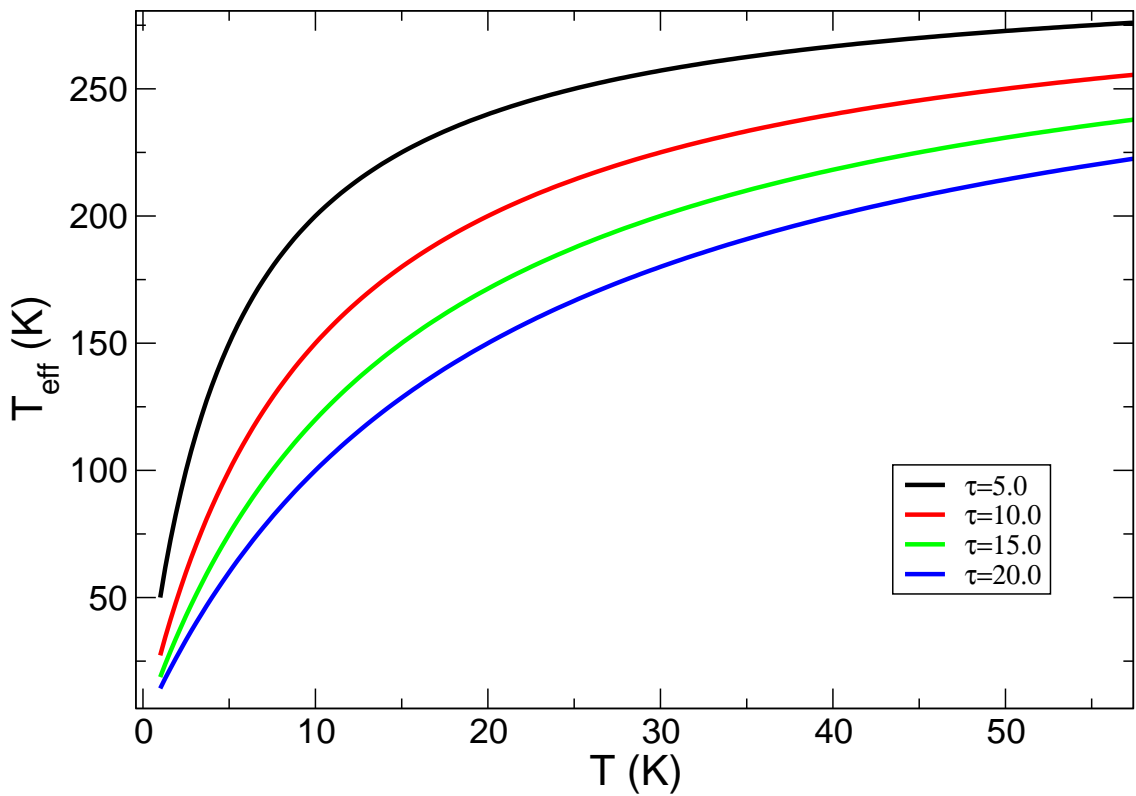}
\caption{Computed behaviour of the $T_{eff}$ parameter as a function of the physical temperature for a selected set of $\tau$ values within the range of interest in the present study. See main text for further details.}
\label{taueff}

\end{figure}

We can try to better understand the meaning of $T_{eff}$ by plotting its variation with the physical temperature of the reaction bath for a different choice of the parameter $\tau$ as given by the range shown in figure \ref{taueff}. We see that in the low temperature range the $T_{eff}$ indicates that the reactive system  partially heats the non-reactive modes for the reasons discussed before (i.e. the TS spends more time along the MEP). Hence the $T_{eff}$ value has to be higher than the actual physical temperature of the thermal bath formed by the reagent molecules. As one moves to higher temperatures, however, we see that $T_{eff}$ essentially coincides with the physical temperature of the reaction as the reaction progress correctly follows the MEP trajectory and the process reaches a pseudo-canonical situation, matching its energy with the Boltzmann's value at room temperatures.

The results gathered by the numerical test of figure \ref{taueff} indicate therefore that the parameter $\tau$ of eq. (6) basically controls the degree of IVR reached by the reactive system: small values of the parameters go with an efficient energy exchanges between reactive and non-reactive modes, which is typical of the lower temperatures for the present exothermic reaction. On the other hand, larger $\tau$ values are associated with an inefficient energy exchange of the reactants with the non-reactive modes (i.e. with the residual degrees of freedom of the reagents), a feature associated here with the near-thermal temperatures where the agreement of our calculations with the experiments do not require the scaling implied by introducing the parameter $\tau$ in eq. (6). Thus, an efficient energy dissipation of the excess reaction energy with the  non-reactive modes (associated with a choice of a smallish value of $\tau < 5$ K) forces the computed rates to grow as the temperature decreases: the actual TS from the VTST approach has to remain 'hot' while the non-reactive modes heat with decreasing temperature. As the temperature increases and becomes closer to the thermal values, then the $T_{eff}$ reaches the room temperature so that it forces the computed rates to decrease with the T of the reagents since now the molecular partion functions of the reagents are at their maximum value while the Q of the TS increases very slowly. Finally, the presence of only a partial IVR effects of the TS is described by the intermediate range of $\tau$ values: $5 {\rm K} <\tau \le 8 {\rm K}$. The computed rates first increase with temperature as the latter decreases from the thermal value, but then begin to decrease with T down to the lower values which are shown by figure \ref{tauscan}. It follows the experimental behaviour, thus indicating that at the lowest temperatures accessed by the  experiments the TS of the exothermic reaction only partially manages to be accessed by the very cold reagents, thereby reducing the probability that the reaction would occur, in agreement with what the experiments have shown.

\section{Present conclusions}

The work presented in the sections above has analyzed several computational aspects of the title reaction, which is of potential interest for the chemical network of N-containing species in molecular clouds and circumstellar envelopes. We have computed both the exothermic reaction from the NH$_{2}^{-}$ + H$_{2}$ reagents, which we have called the FR reaction path, and the slightly endothermic reaction from the NH$_{3}$ + H$^{-}$ as reagents, which we have called the RR reaction path. Through an analysis of its ab initio features, we have characterized two different regions of intermediate structures before the actual formation of a transition state,
the latter being below the reference energy of the reagents for the exothermic process and only slightly above the reagents for the RR mechanism. One of the intermediate structures, labelled Min$_1$ in our work, is a sort of pseudo-minimum associated with the H$_{2}$ molecular partner bound to the NH$_{2}$ molecule, on which most of the excess charge of the anionic complex still resides. The second minimum structure, labelled here the Min$_2$ complex, turns out to be a true minimum and describes a structure located in the product region for the FR path: the excess charge has now moved almost entirely onto one of the H atoms, forming a complex between H$^{-}$ and the NH$_{3}$, nearly neutral, molecule.

The existence of an embedded transition state structure along a barrierless path for the exothermic reaction allowed us to model the latter by using the VTST approach, already employed earlier to study ionic reactions in the ISM environments \cite{key-26,key-37}. The current calculations show very good agreement between our VTST rates and the existing experimental data at room temperature, reported by ref. \cite{key-31}. However, once we extended the VTST approach to lower temperatures down to about 8 K, the calculations produced much smaller rates than those given by the experiment of Otto {\it et al.} \cite{key-32}. The VTST-modelled rates, which are treated as following the MEP when going down to sub-thermal conditions, are too rapidly decreasing with the lowering of the reaction temperature. We then suggested a temperature-dependent scaling procedure which links the room temperature rates to their experimental values but increases the efficiency of the TS energy dissipation process into the non-reactive degrees of freedom as T decreases. We thus found that, with such a scaling, we can reproduce very accurately the experimental rates for the FR exothermic reaction down to the lowest measured T values around 8 K. The slightly endothermic RR process, however, is not greatly modified by the present scaling and still indicates a rapid decrease of the endothermic rates down to very low temperatures. No experimental data are available for the RR process for comparison with our computed results.

In conclusion, we have shown that it is possible to modify in the low-T regimes the VTST approach and carry out rate calculations for barrierless, exothermic reactions by introducing a temperature-dependent factor defined in eqs. (6) and (9) as the factor $\tau$, within the reagents' ZPE parameters. This parameter is further shown to regulate the efficiency of the low-T energy dissipation of the TS structures within the increasingly colder reagents which contain the non-reactive degrees of freedom that are coupled with the TS during the reactive encounters as the physical temperature goes down well below room temperature. The role of this parameter also explains why, as T becomes smaller, the experimental and computed rates resume behaving as decreasing with temperature. Our analytic fitting of the final rates provides here realistic extimates for the FR reaction rates at low temperatures. Our new data could then be directly employed when modeling the evolution of the title molecules within larger astrophysical chemical networks involving nitrogen-containing species \cite{a1}. At the same time, the present study  also provides a fairly detailed structural description of the reaction paths and of the energetics involved when going from reagents to products along both the reactions of the title systems.

\section*{Acknowledgments}

F.A.G. and R.W. acknowledge the support by the Austrian Science Fund (FWF),
Project No. 29558-N36.

\providecommand{\latin}[1]{#1}
\makeatletter
\providecommand{\doi}
  {\begingroup\let\do\@makeother\dospecials
  \catcode`\{=1 \catcode`\}=2 \doi@aux}
\providecommand{\doi@aux}[1]{\endgroup\texttt{#1}}
\makeatother
\providecommand*\mcitethebibliography{\thebibliography}
\csname @ifundefined\endcsname{endmcitethebibliography}
  {\let\endmcitethebibliography\endthebibliography}{}

\end{document}